\documentclass{aa}  
\usepackage{graphicx}
\usepackage{txfonts}
\usepackage{hyperref}
\usepackage[bottom]{footmisc}
\usepackage{natbib}
\bibpunct{(}{)}{;}{a}{}{,}
\usepackage{subfigure}
\usepackage{multirow} 

\usepackage{lscape}
\usepackage{tabularx}
\usepackage{float}
\hypersetup{colorlinks=true, allcolors=blue}
\defcitealias{Edenhofer2024}{E24}

\usepackage{placeins}

\begin{document}

   \title{The superclouds of the local Milky Way}


   \author{
   Lilly A. Kormann\inst{\ref{univie}},
   Jo\~ao Alves \inst{\ref{univie}},
   Michelangelo Pantaleoni Gonz\'alez \inst{\ref{univie}},\\
   Cameren Swiggum \inst{\ref{univie}},
   Torsten~A.~En{\ss}lin \inst{\ref{mpa}, \ref{lmu}, \ref{zfa}, \ref{cluster_o}}, 
   Gordian Edenhofer\inst{\ref{mpa}}
   }
   \authorrunning{Kormann, L. A., et al.}

   \institute{
    University of Vienna, Department of Astrophysics, T\"urkenschanzstrasse 17, 1180 Wien, Austria\label{univie}
        \and
    Max-Planck Institut für Astrophysik, Karl-Schwarzschild-Str. 1, 85748 Garching, Germany\label{mpa}
        \and
    Ludwig-Maximilians-Universität M\"unchen (LMU), Geschwister-Scholl-Platz 1, 80539 M\"unchen, Germany\label{lmu}
        \and
    Deutsches Zentrum f{\"u}r Astrophysik, Postplatz 1, 02826 G{\"o}rlitz, Germany\label{zfa}
        \and
    Excellence Cluster ORIGINS, Boltzmannstr. 2, 85748 Garching, Germany\label{cluster_o}
             }

   \date{Received 17 July 2025 / Accepted 3 December 2025}

  \abstract
   {Recent 3D dust maps of the local Milky Way are revolutionizing our understanding of the Sun’s Galactic neighborhood, providing much needed insight into the large-scale organization of the interstellar medium. Focusing on the largest scales in \textit{Gaia}-based 3D dust maps, we find a pattern of seven highly elongated, mostly parallel structures in the local $\sim 5\,\mathrm{kpc}^2$, five of which were previously unknown. These structures show pitch angles of $33.5 \pm 4.0 \degr$ and masses ranging from $10^5$ to $10^6$ M$_\odot$. We refer to these structures as superclouds. Nearly all known star-forming regions in the solar neighborhood lie within the superclouds, primarily along their central axes, supporting the idea that they act as gas reservoirs for the formation of giant molecular clouds. All but one of the seven superclouds show an underlying undulation, indicating that this is not a property unique to the Radcliffe Wave. We find that while the superclouds have linear masses that vary by about a factor of 4, their volume densities only vary by about 10\%. This suggests that superclouds self-regulate their physical sizes and internal structure to maintain pressure equilibrium with their environment. These findings establish a new framework for understanding how large-scale Galactic structures shape the conditions for star formation in the solar vicinity, and likely in galaxies like the Milky Way.
   }

   \maketitle
%

\section{Introduction}

The interstellar medium (ISM) of the Milky Way is a complex and structured environment characterized by features spanning a wide range of spatial scales. Among these, superclouds represent some of the most massive components. The concept of a supercloud first appeared in the seminal work of \cite{elmegreen1983}, who highlighted the regular distribution of star complexes and their associated neutral atomic hydrogen (H\textsc{i}) superclouds along galactic spiral arms in nearby galaxies. They proposed that these large-scale structures, with typical masses between $10^6$ and $10^7$ M$_\odot$ and lengths on the order of a few 100 pc to 1 kpc, originate from gravitational or magneto-gravitational instabilities in spiral arms. This provided a link between galactic dynamics and the formation of ISM structures, motivating further investigations into the processes that govern the assembly of massive gas concentrations in galaxies. 

Expanding on this groundwork, \citet{Elmegreen1987} identified H\textsc{i} superclouds in the inner Galaxy and noted their spatial correlation with giant molecular cloud (GMC) complexes. Similarly, \citet{Dame1986}, using carbon monoxide (CO) observations, revealed large molecular complexes that are embedded within superclouds. \cite{Grabelsky1987} conducted a large-scale CO survey of the Carina arm, demonstrating that molecular clouds are additional Population I tracers for this spiral arm, and established a strong link between their distribution and kinematics with those of neutral atomic hydrogen. \cite{Lada1988} presented coordinated H\textsc{i} and CO observations of M31 at the same spatial and spectral resolution to resolve individual GMCs. They found that the M31 GMCs are similar to the Milky Way GMCs and are embedded in a large H\textsc{i} spiral arm filamentary feature (see their Fig. 1). 

Recently, \cite{park2023} identified several H\textsc{i} superclouds in the Carina arm associated with H\textsc{ii} regions, indicating ongoing star formation within these larger atomic structures, presumably occurring within the embedded molecular clouds. Their analysis of the star formation rate surface density also suggests a strong correlation with the molecular gas surface density, but little correlation with the atomic hydrogen surface density, highlighting the critical role of the molecular phase in the process of stellar birth. 

Subsequent theoretical work on the origin of large-scale ISM structures in galaxies was conducted by \cite{Franco2002}. They explored the Parker instability in the context of spiral arms, demonstrating via magnetohydrodynamic simulations that it can lead to the formation of large-scale gas condensations and thus, potentially, superclouds. These structures are distributed with a characteristic wavelength along the arm and are preferentially positioned alternately above and below the galactic midplane. \cite{Kim2001,kim2002} further emphasized the roles of self-gravity and magnetic fields as the key to gravitational runaway, and explored the formation of large-scale spiral arm features such as spurs and their fragmentation into GMCs. There is a general consensus from observations and theory that favors a ``top-down'' (instability) rather than ``bottom-up'' (coagulation) process for the formation of at least the most massive clouds ($\gtrsim 10^5 - 10^6$ M$_\odot$), where most of the molecular material and star formation are found \citep{kim2003,mckee2007}.

Despite their recognition on galactic scales, the properties of superclouds in the immediate solar neighborhood remain poorly constrained. Several factors contribute to this knowledge gap. 
The isolation of large-scale complexes from the diffuse and extended background of the ISM is a challenge. Identifying individual clouds in H\textsc{i} is especially difficult due to its widespread and diffuse nature. Combined with line-of-sight overlap, this complicates the establishment of clear structure boundaries. While CO serves as an effective tracer of dense molecular gas, it fails to capture the more diffuse components of the ISM, limiting our ability to trace the full extent of large-scale gaseous structures. Moreover, kinematic distance ambiguities, particularly toward the inner Galaxy, have limited the effectiveness of spectroscopic surveys \citep[e.g.,][]{Dame1986,Elmegreen1987}. 
The possibility that the Sun is embedded within the remnants of an ancient supercloud, as suggested by \citet{Olano2016}, further complicates identification. Finally, the definition of a supercloud, whether it emphasizes atomic (H\textsc{i}) or molecular gas (as traced by CO), varies between studies, making it difficult to establish a uniform and comprehensive survey of such structures in the local Galactic environment. 

The recent discovery of the Radcliffe Wave, a 2.7 kiloparsec-long chain of dense gas clouds in the solar neighborhood \citep{alves2020}, has renewed interest in the study of large-scale structures in the ISM. This remarkable structure exhibits an unexpected wavelike shape with a maximum amplitude of about 160 pc and had remained elusive despite its size and close proximity to the Sun (within 300 pc at its closest point). Like a supercloud, the Radcliffe Wave is composed of several massive star-forming regions (SFRs) embedded in well-known GMCs and connected by less dense (likely atomic) gas. Its surprising oscillation about the Galactic plane and radial drift away from the Galactic Center \citep{konietzka2024} provides a nearby example of a large-scale Galactic-size gaseous feature but raises questions about the uniqueness of this particular gas arrangement.

 To address these challenges, we present a study of newly identified superclouds in the local $\sim 5\,\mathrm{kpc}^2$ of the Milky Way. Our approach takes advantage of recent advances in 3D dust mapping from large-scale photometric surveys, particularly those incorporating ESA \textit{Gaia} data \citep[e.g.,][]{Green2019,Lallement2019,Chen2019,Leike2020,Edenhofer2024}. By analyzing the 3D distribution of interstellar dust, an effective tracer of the total gas column density \citep[e.g.,][]{lada1994,alves1999,lombardi2006}, we identified regions of enhanced density and find a well-defined pattern of large-scale cloud structures. This method offers a complementary perspective to traditional kinematic analyses and avoids many of the longstanding issues, such as velocity crowding, line-of-sight confusion, and distance uncertainties.

In this study we segmented the 3D dust map of \cite{Edenhofer2024} and constructed the first catalog of superclouds within the local kiloparsec of the Galaxy. We characterized their physical properties, including filling factor, length, width, height, orientation, and mass, on the basis of the 3D dust distribution, and find a pattern of seven highly elongated, mostly parallel kiloparsec-long superclouds in the local Milky Way. We examined the connection between these superclouds and the broader ISM in the solar neighborhood and explored potential associations with known stellar populations and other ISM components. Overall, our results provide new insights into the distribution and properties of the most massive interstellar structures in the local Galaxy, with implications for Galactic structure and the early stages of GMC formation.

\begin{figure*}[t]
    \centering
    \begin{minipage}[b]{0.49\textwidth}
        \centering
        \includegraphics[width=\textwidth]{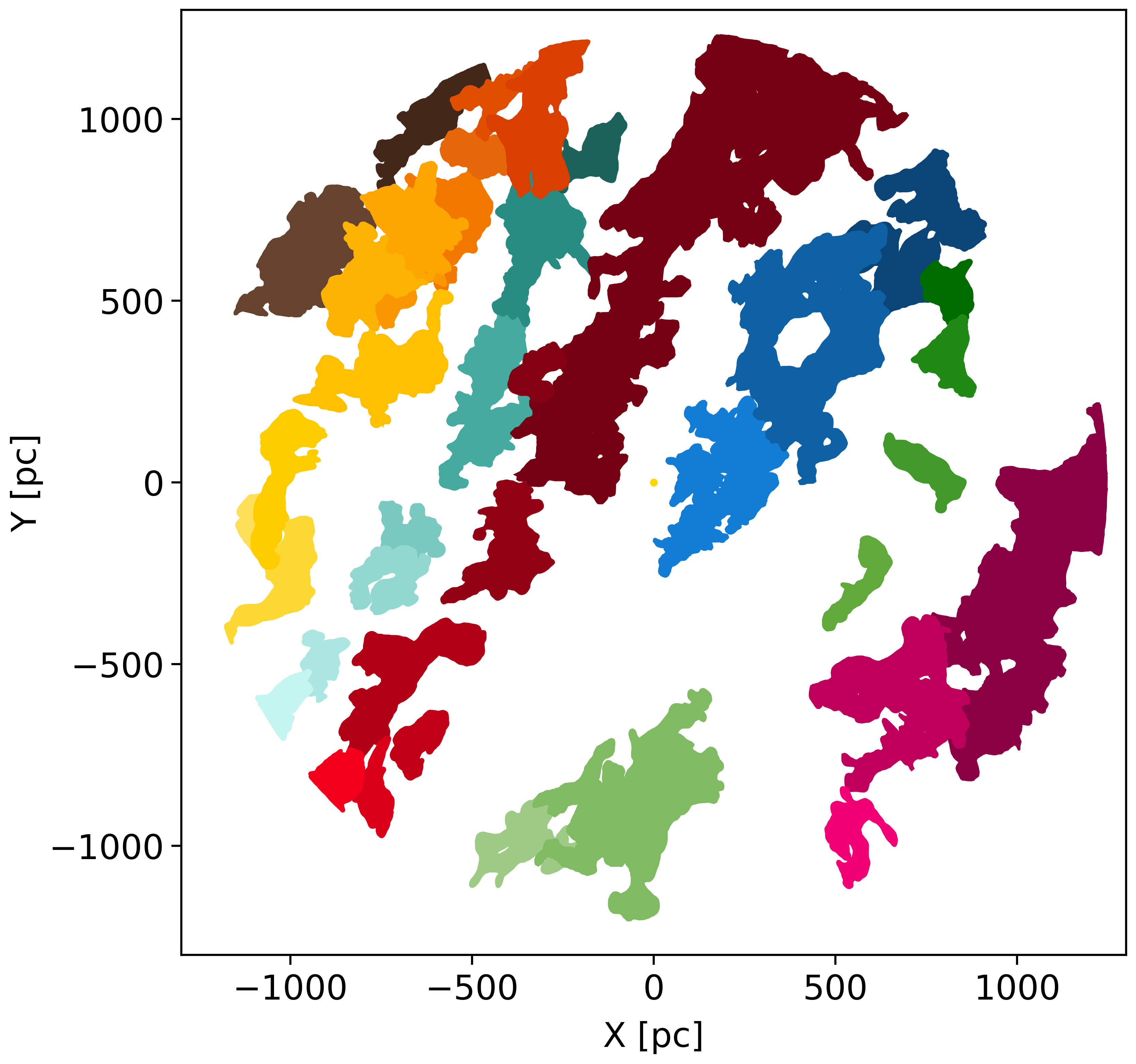}
    \end{minipage}%
    \hfill
    \begin{minipage}[b]{0.49\textwidth}
        \centering
        \includegraphics[width=\textwidth]{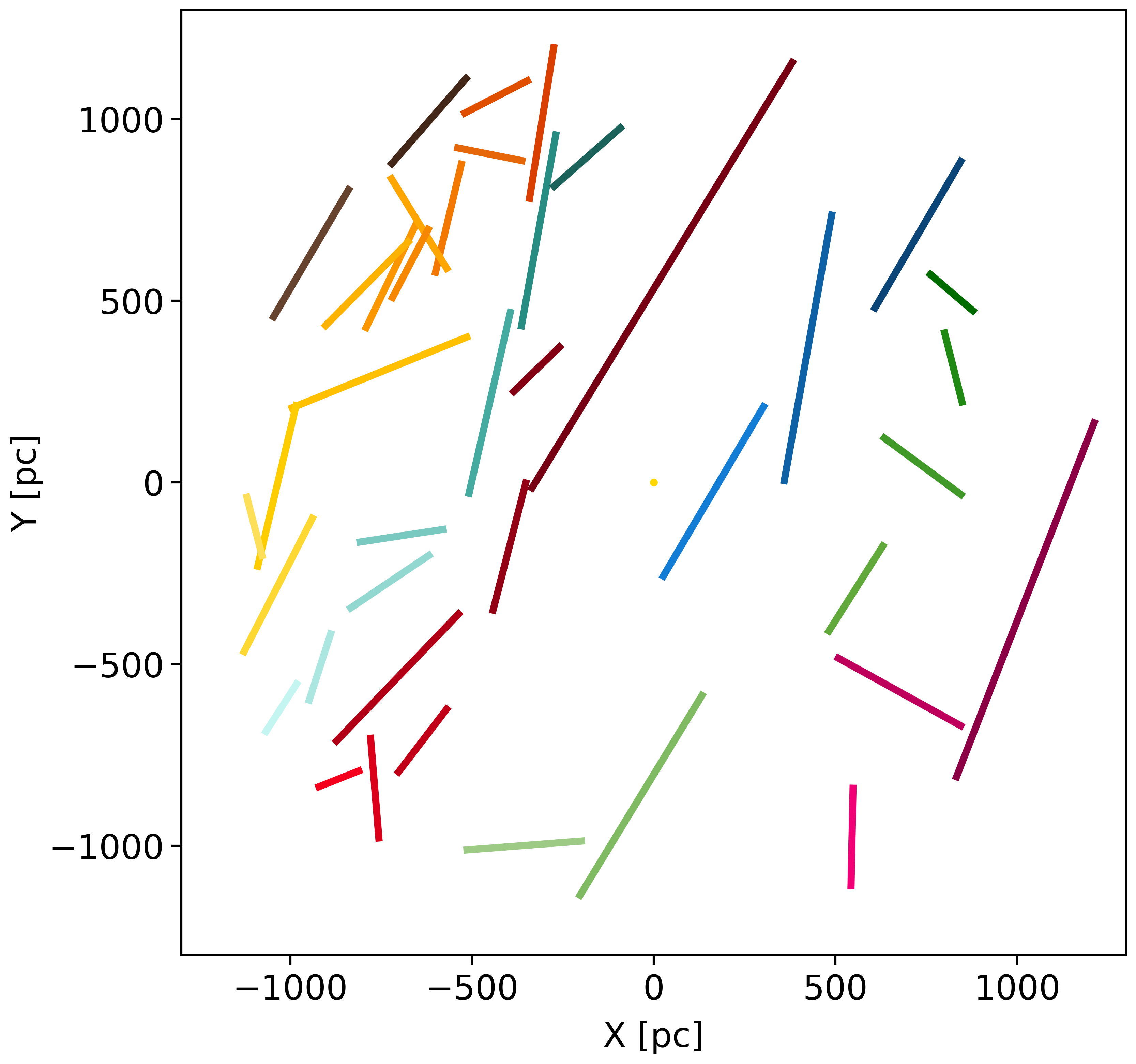} 
    \end{minipage}
    \caption{Face-on view of the 40 HOP clouds (left) and their corresponding orientation (right). Each cloud is assigned a distinct color, and the length of the line corresponds to the calculated length of the cloud. The Sun is represented by the yellow dot in the center; the Galactic Center is located to the right. An interactive 3D version of the 40 HOP clouds is available \href{https://lillykormann.github.io/superclouds_figures/hop_clouds/hop_clouds_surface.html}{here}.}
        \label{fig:structures_general}
\end{figure*}

\section{Data}

We used the posterior mean of the 3D Cartesian grid dust map interpolated by \citet[hereafter \citetalias{Edenhofer2024}]{Edenhofer2024}, which extends out to 1.25 kpc from the Sun. The map is based on stellar distance and extinction estimates from \cite{zhang2023}. To obtain the reconstruction, the logarithm of the dust extinction is modeled using a Gaussian process, implemented in the \texttt{NIFTy.re} framework \citep{edenhofer2024_nifty}, which incorporates smoothness priors to counteract the "fingers-of-god" effect. This is enabled by a new generative method called iterative charted refinement \citep[][]{edenhofer2022_icr}, which efficiently models Gaussian processes on nearly arbitrarily spaced voxels, iteratively refining the map's resolution until a target resolution is reached. The original reconstruction is performed on logarithmically spaced HEALPix spheres with decreasing angular resolution with distance from the Sun. 

For the posterior mean of the Cartesian interpolated map, we have a resolution of $2^3$ pc$^3$, with the data given as unitless extinction values. To convert the data to hydrogen volume density, we applied 
    \begin{equation}
        n_i = 1653\,[\text{cm}^{-3}] \cdot A_i, \label{eq:dust_density}
    \end{equation}
where $A_i$ is the extinction value in the $i$-th voxel, correlated with hydrogen volume density $n_i$ in units of cm$^{-3}$ via a factor of 1653 cm$^{-3}$. The derivation of the conversion formula is detailed in \cite{oneill2024}. By applying this conversion, we obtained a 3D hydrogen volume density distribution map.

\section{Methods}
\label{sec:methods}

  To analyze and define the large-scale structures in the dust map, four different methods are tested: the scikit-image package \citep{scikitimage}; dendrograms via the astrodendro package \citep{astrodendro}; HOP \citep{colman2024}, a package developed for the extraction of molecular clouds in numerical simulations; and a deep learning approach modified at the Research Center Data Science of the University of Vienna (DS/Univie). Ultimately, the HOP algorithm was selected as the best fit for this project. A discussion of the alternative methods and the reasons for their rejection can be found in Appendix \ref{ap:methods}.

    \subsection{HOP structure finder}
    
    The structure finding algorithm by \cite{colman2024} is based on the HOP clump finder of \cite{eisenstein1998}. HOP was originally designed to identify groups in N-body simulations, and was later adapted by \cite{colman2024} to extract and compare molecular clouds in numerical simulations. 
    
    HOP depends on an externally set density threshold and peak density for the data segmentation, which precludes a solely data-driven approach. However, it works remarkably well given the large size of the dust map compared to the other methods, and the four segmentation parameters are well suited for the problem. 

    \subsubsection{Segmentation stages}
    HOP goes through two main stages to identify structures, the initial segmentation and the structure definition. Before considering any input parameters, the algorithm checks for density peaks around local volume density maxima, resulting in a preliminary segmentation. 
    
    To define a structure in the second stage, the algorithm takes four parameters into account:
    \texttt{density\_threshold}, \texttt{mincells}, \texttt{density\_peak}, and \texttt{density\_saddle}.
    The minimum density required for a voxel to be considered as part of a structure is set by \texttt{density\_threshold}, the initial density cut. All voxels below this threshold are discarded. The parameter \texttt{mincells} defines the minimum number of cells for an individual structure, structures unable to meet this requirement are discarded or merged with a neighboring structure. To be considered significant, a peak from the initial segmentation must have a minimum density of \texttt{density\_peak}. Lastly, \texttt{density\_saddle} defines the maximum density threshold for two structures to be seen as individual structures. If two structures have neighboring voxels, HOP identifies boundary pairs, consisting of a border voxel of structure A and structure B. Each boundary pair has a boundary density value that is calculated as the average of their densities. The largest boundary density value per neighboring structures defines the \texttt{density\_saddle} value. If the \texttt{density\_saddle} value is above the set threshold, the structures are merged and seen as substructures of the main structure. Once the segmentation is done, the field \texttt{structure\_id} is added to the data, assigning each voxel to a structure. Voxels with label -1 belong to no structure. 
    
    HOP includes a possible third stage to calculate multiple structure properties depending on the available data. This stage is skipped, as the only available data is the density. 

    \subsubsection{Parameter selection}
    
    To find large-scale structures in the \citetalias{Edenhofer2024} dust map, the following HOP parameters were selected after multiple test runs: \texttt{density\_threshold} = 1.3 cm$^{-3}$, \texttt{density\_saddle} = 1.3 cm$^{-3}$, \texttt{density\_peak} = 6 cm$^{-3}$, and \texttt{mincells} = 60.000. 
    
    The \texttt{density\_threshold} was chosen to recover and distinguish the largest present structures, while avoiding merging across the entire map due to low-density connections. Multiple parameter tests with \texttt{density\_threshold} values between 1 cm$^{-3}$ and 5 cm$^{-3}$ have shown that the exact value of the threshold does not significantly impact the general trend of the recovered clouds. While the threshold itself is below the resolution limit at which individual low-density structures can be reliably resolved, the integrated extinction measurements constrain the dust distribution well enough to identify large-scale structures with lower densities, even if their boundaries cannot be precisely traced on a parsec scale. At higher values, for both \texttt{density\_threshold} and \texttt{density\_peak}, structures start to disappear, and only the densest regions remain. 
    The parameter \texttt{density\_saddle} controls whether structures are separated or merged based on the density of their connecting points. In this work, it was set equal to \texttt{density\_threshold} to support the identification of large-scale structures. 
    The purpose of \texttt{density\_saddle} is to prevent the connection of structures via thin, low-density bridges, which is not an apparent problem in the final segmentation. Increasing \texttt{density\_saddle} results in a more fragmented segmentation, breaking apart coherent clouds, and excluding otherwise present clouds of interest. 
    Lastly, \texttt{mincells} is set to filter out smaller structures that might meet the other criteria, but are not of interest in this study.
    
    \cite{eisenstein1998} demonstrated that the density threshold parameter has the greatest influence on the resulting group properties. Changing the initial density cut redefines the outer boundaries of the structures and determines how many voxels are included in the segmentation. In contrast, the remaining parameters primarily influence the internal structure and merging behavior.

\subsection{Structure properties}

    \subsubsection{Mass calculation}
    
    For the mass calculation, we used 
    \begin{equation}
        M = 1.37 \,m_p \sum_i n_i \,\text{d}v_i \label{eq:mass_structures}
    \end{equation}
    from \cite{oneill2024}. Here, $m_p$ is the proton mass, the factor 1.37 is used to convert from hydrogen mass to total mass \citep{heiderman2010}, and $\text{d}v_i$ denotes the volume of each voxel. For the Cartesian map, the volume of each voxel is $2^3$ pc$^3$.

    \subsubsection{Shape definition}
    
    To describe the 3D shape, we applied principal component analysis (PCA; \citealt{jolliffee1986pca}) to each individual structure. PCA identifies the direction of largest variance within a dataset, determining the primary orientation of the structure. In the case of 3D data, PCA extracts three mutually orthogonal components, creating a new set of basis vectors that aligns with the intrinsic shape of the structure. The first principal component (PC1) describes the direction of the largest variance and the length of the structure. The second principal component (PC2) describes the direction of the largest variance orthogonal to PC1 and the width of the structure. The third principal component (PC3) describes the direction of the largest variance orthogonal to PC1 and PC2, and the height of the structure. 
    
     We determined the structure axis lengths by projecting the data onto the principal components and measuring the extent between the minimum and maximum, thereby providing an upper bound. For the length, this is a good approximation, but due to the irregular shape of the clouds, defining the height and width leads to a large overestimation in the majority of the cases. 

    \subsubsection{Orientation calculation}
    \label{sec:orientation}
    
    The structure orientation is determined by the angle between PC1 and a chosen reference axis. This angle is calculated by applying the arc cosine to the absolute value of the cosine similarity between the two vectors, and then converting the result to degrees. For consistency and easier comparison, all structures are defined as oriented in the positive direction relative to the reference axis, with angles measured clockwise from it. The angle between PC1 and the Y-axis (face-on) is defined as $\alpha$, while $\beta$ represents the angle between PC1 and the Z-axis (edge-on). Circular statistics are applied to calculate the mean and standard deviation of all angles $\alpha$ and $\beta$. The angles are weighted by mass to consider each structure's relative importance.

    \subsubsection{Undulation fitting}
    \label{methods:undulation_fitting}
    
    We fit the projected superclouds in the Z/Y$^\prime$ plane with a damped sinusoidal of the form
        \begin{equation}
        f(z) = a\, e^{\epsilon_a z} \, \sin\left(\omega \cdot e^{\epsilon_\omega z} \, z + \phi\right) + d,
        \label{eq:damp_sin}
        \end{equation} 
    where $z$ corresponds to the projected Y$^\prime$-coordinate, obtained by rotating the superclouds counterclockwise by their average angle $\alpha$ to lie in the plane, and $f(z)$ gives the corresponding Z-position. The parameter $a$ is the amplitude, $\omega$ the angular frequency, $\phi$ the phase shift, and $d$ the vertical offset.  The function includes an exponential modulation of both amplitude and frequency: $\epsilon_a$ describes the exponential growth or decay of the amplitude along the wave, while $\epsilon_\omega$ describes a linear change in frequency. Positive values for $\epsilon_a$ indicate a growth in amplitude over distance, while negative values indicate damping. For $\epsilon_\omega$, a positive value indicates a growth in frequency, resulting in a decrease in wavelength via the relation $\lambda = 2\pi / \omega$, where $\lambda$ is the wavelength and $\omega$ the angular frequency. 
    
    The fitting approach is simplified to give an initial characterization of the underlying undulation and should not be interpreted as implying a physical oscillation.

\section{Results}

\subsection{HOP cloud catalog}

HOP identifies 40 clouds in the \citetalias{Edenhofer2024} dust map, visualized in the left panel of Fig.~\ref{fig:structures_general}. The cloud masses range from $3.8 \times 10^4$ to $2.8 \times 10^6$ M$_\odot$, a selection of the calculated cloud properties including the mass and orientation can be found in Table \ref{tab:cloud_properties}. All clouds with their corresponding ID numbers are visualized in Fig. \ref{fig:structures_full_with_id}.

    \begin{figure}
        \centering
        \includegraphics[width = 0.48\textwidth]{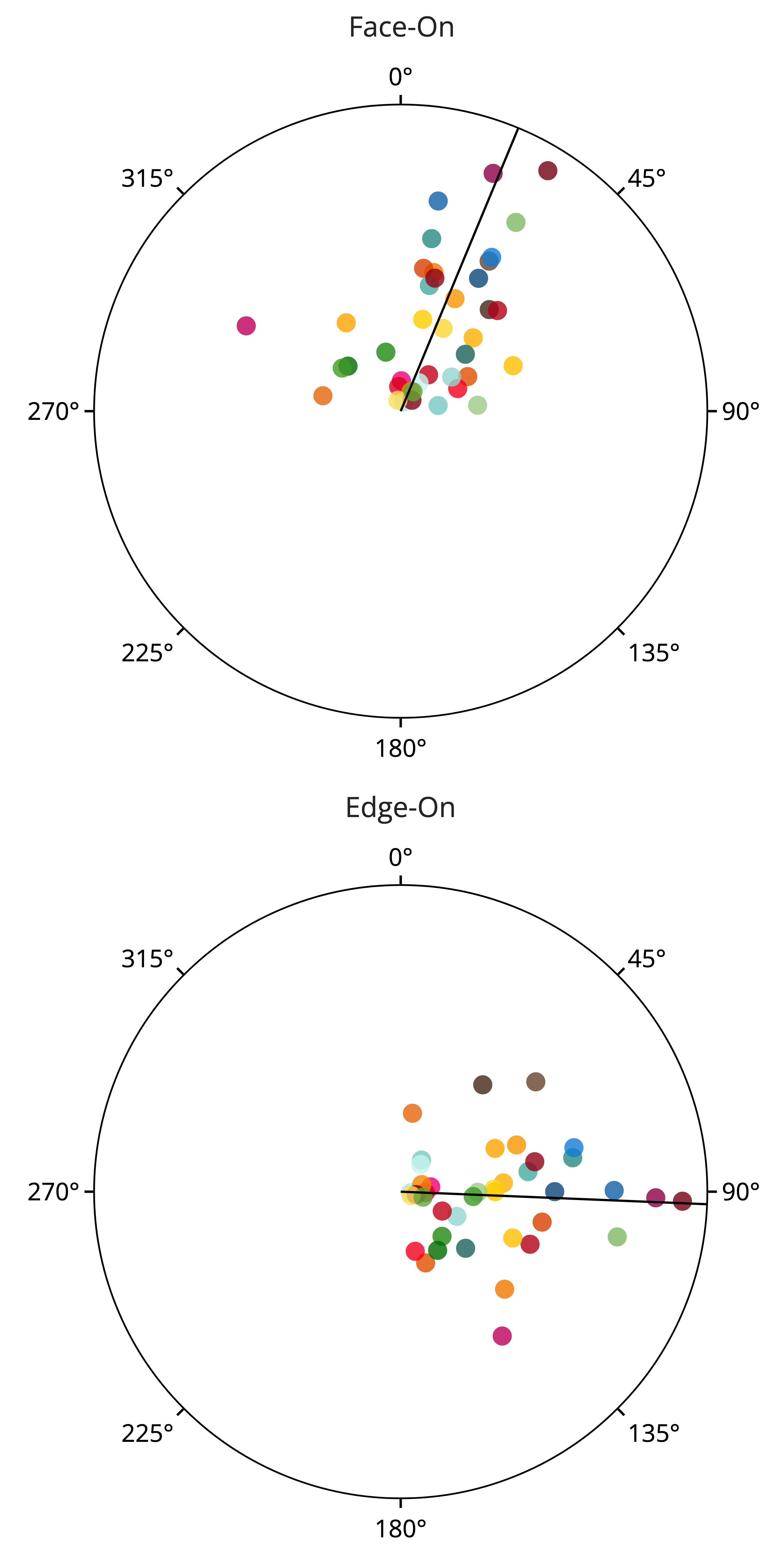}
        \caption{Orientation distribution of the 40 HOP clouds. Colors indicate the same clouds as in Fig. \ref{fig:structures_general}, the distance from the center reflects the cloud mass, and the solid line represents the mean orientation. Top: Face-on view from the North Galactic Pole, with an average angle of $\alpha = 22.6 \pm 25.5 \degr$. Bottom: Edge-on view from the Galactic Center, with an average angle of $\beta = 92.3 \pm 23.4 \degr$.}
        \label{fig:structures_angle_distribution}
    \end{figure}
    
In the right panel of Fig.~\ref{fig:structures_general} we plot the face-on orientation for each cloud, defined by their angle $\alpha$. The distribution of all measured angles $\alpha$ and $\beta$ is visualized in Fig.~\ref{fig:structures_angle_distribution}. More massive clouds show a preferred orientation of approximately 30$\degr$, whereas smaller clouds tend to deviate, especially in the upper left region of the orientation plot. The average mass-weighted orientation for all clouds is $\alpha = 22.6 \pm 25.5 \degr$ face-on, and $\beta = 92.3 \pm 23.4 \degr$ edge-on. In general, we see a quasi-parallel alignment, which is highlighted by a color gradient in the visualization. This trend is most prominent in the turquoise, red, and blue cloud groups. Some clouds following the trend can also be seen in the orange, green, and pink groups, but overall their orientation is more chaotic (right panel of Fig.~\ref{fig:structures_general}). The brown clouds in the upper left corner are not considered to be part of the neighboring yellow group.

\begin{figure}
        \centering
        \includegraphics[width = 0.49\textwidth]{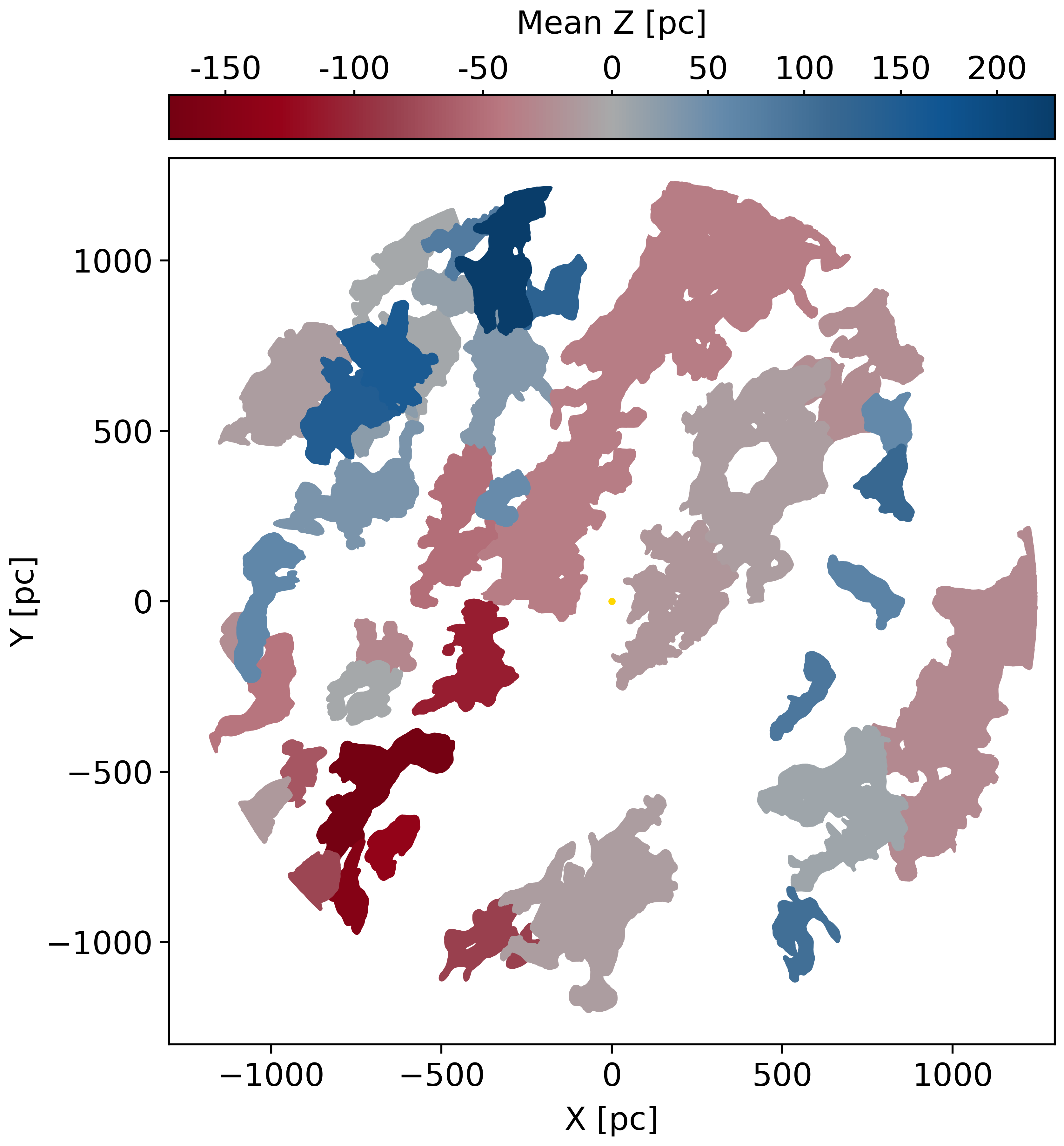}
        \caption{Face-on view of the HOP clouds color-coded by their mean altitude in Z. The Sun is represented by the yellow dot in the center, and the Galactic Center is located to the right.}
        \label{fig:structures_cc_meanz}
    \end{figure}

The HOP clouds, color-coded by their mean altitude in Z, can be seen in Fig.~\ref{fig:structures_cc_meanz}. Larger and more massive clouds are primarily located near the Galactic plane, while smaller clouds are also found at both higher and lower altitudes. Considering the proposed quasi-parallel cloud alignment, underlined by the color gradient in Fig.~\ref{fig:structures_general}, the color-coding reveals a pattern, where elevated clouds are followed by a downward dip, implying an underlying undulation.

\subsection{Supercloud catalog}

\begin{figure*}[htb]
    \centering
    \begin{minipage}[b]{0.49\textwidth}
        \centering
        \includegraphics[width=\textwidth]{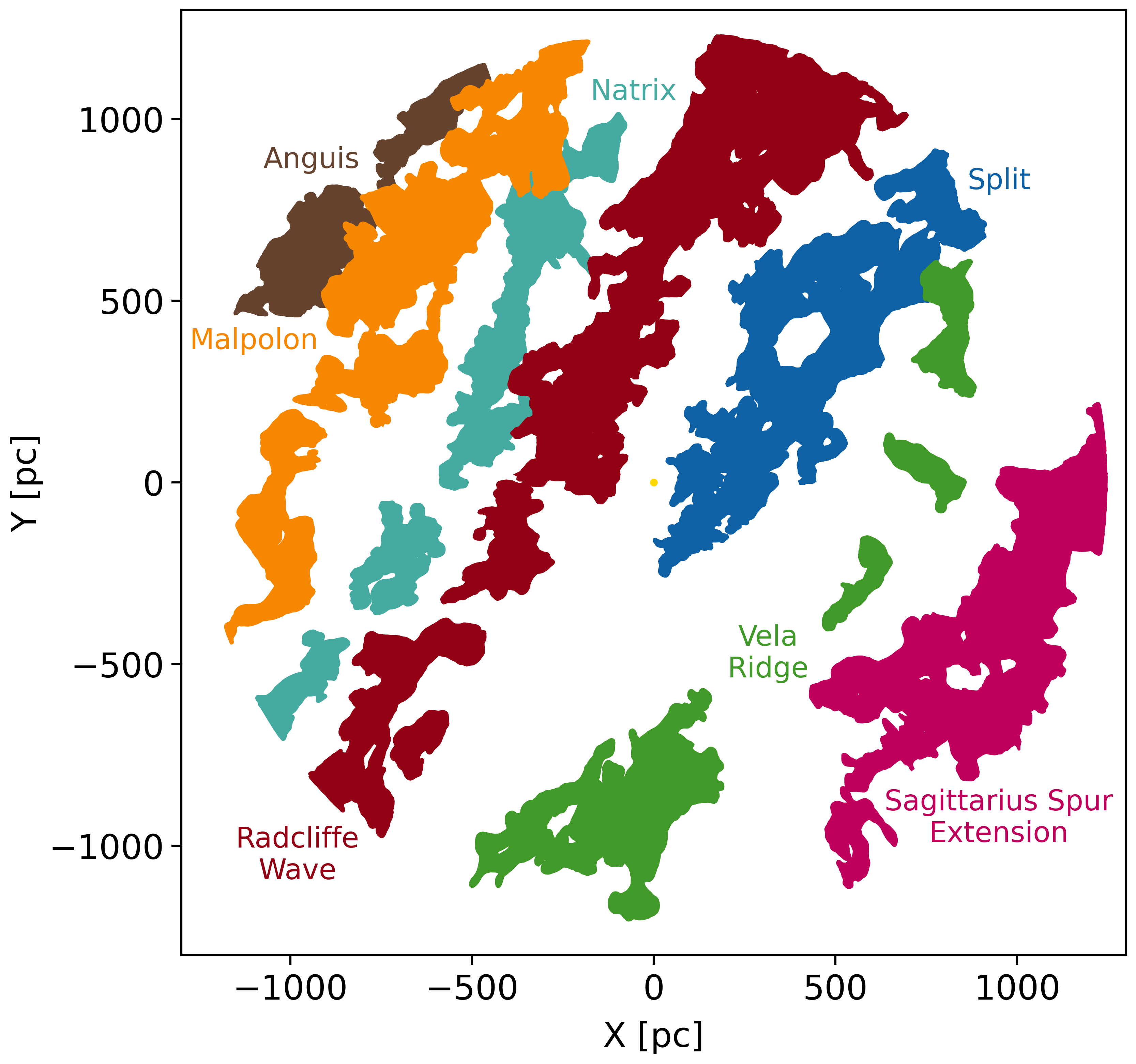}
    \end{minipage}%
    \hfill
    \begin{minipage}[b]{0.49\textwidth}
        \centering
        \includegraphics[width=\textwidth]{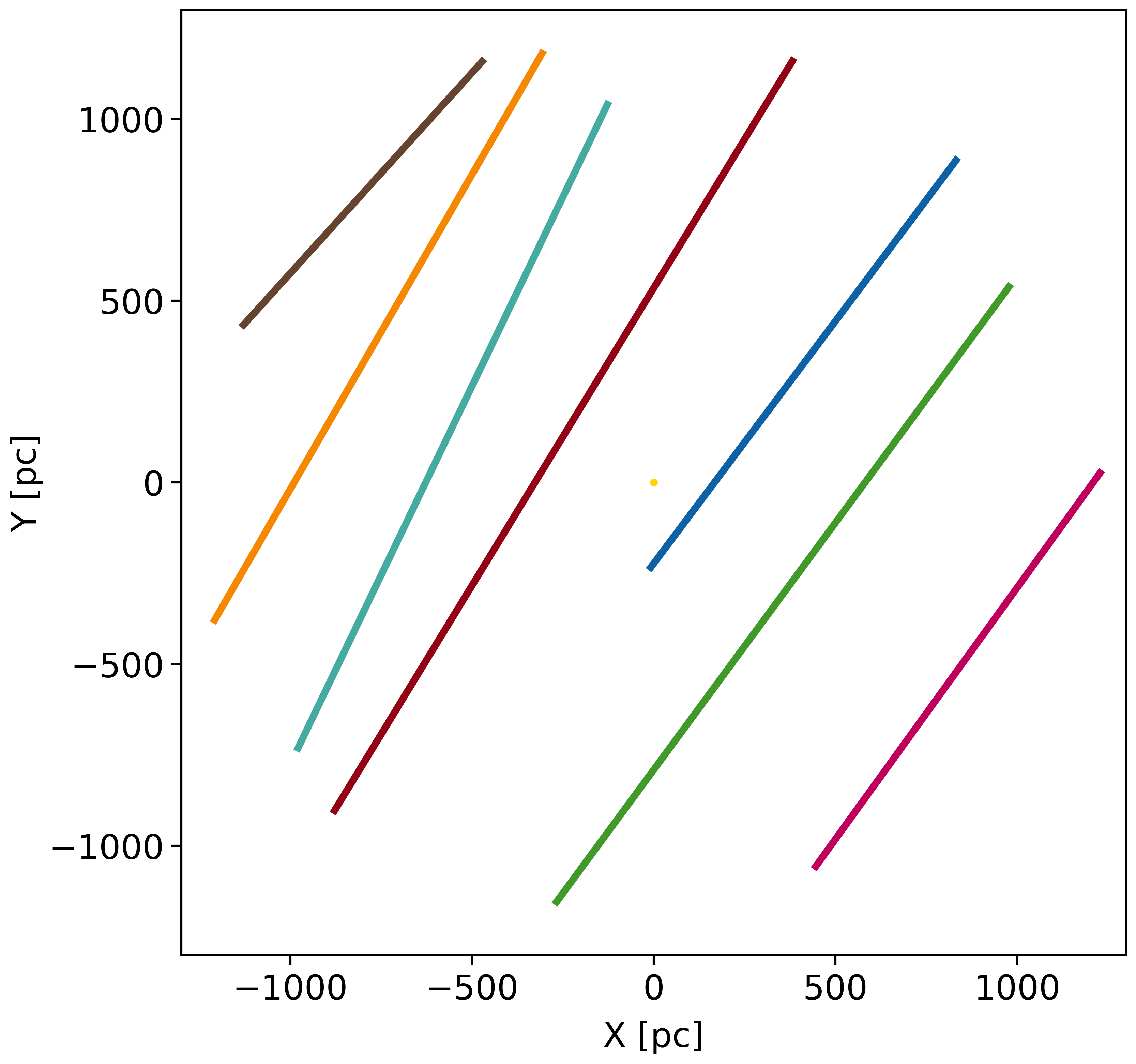}
    \end{minipage}
    \caption{Face-on view of the seven superclouds defined in the local Milky Way (left) and their corresponding orientation (right). Each cloud is assigned a distinct color. The length of the line corresponds to the calculated length of the cloud. The Sun is represented by the yellow dot in the center, and the Galactic Center is located to the right. An interactive 3D version of the seven superclouds is available \href{https://lillykormann.github.io/superclouds_figures/superclouds/superclouds_surface.html}{here}.}
    \label{fig:superclouds_general}
\end{figure*}

Taking the orientation, proximity, and 3D alignment of the HOP clouds into account, we identify seven superclouds, visualized in the left panel of Fig.~\ref{fig:superclouds_general}. The corresponding HOP cloud components are listed in Table~\ref{tab:superclouds_table}, with the supercloud ID numbers visualized in Fig. \ref{fig:superclouds_full_with_id}.

To connect the superclouds, the first criterion is the face-on orientation of the HOP clouds, defined by $\alpha$ and shown in the right panel of Fig.~\ref{fig:structures_general}. The primary motivation for grouping these structures lies in their coherent orientation. 
If adjacent clouds show a similar orientation, defined as differing by no more than 35\degr, they are considered as initial candidates for a supercloud. This threshold is chosen to connect coherent, large-scale structures while avoiding random overlaps.

We then verified the preliminary connections by taking the spatial distribution of the clouds into account. This ensures that the clouds grouped by orientation form a consistent, physical structure. Candidate clouds that share similar orientations, are spatially connected, and appear as a single extended feature are confirmed as parts of the superclouds. Clouds that deviate from the group orientation, yet still show a spatial connection, are also included as candidate clouds. This case of spatial connectivity is only applied when neighboring clouds clearly form a continuous feature.

It is likely that external forces have played a role in the fragmentation of the superclouds during their lifetime. Clouds showing no connection to the supercloud candidates are verified concerning their general 3D alignment. If displaced clouds follow the spatial trend of a larger, already verified group, they are accepted candidates based on their 3D alignment. If these clouds cannot be assigned to another candidate group following the mentioned steps, they are confirmed as part of the first suggested supercloud. 

The connection of the superclouds is re-verified by comparing the final segmentation to the parameter test runs with different lower \texttt{mincells} parameters, ensuring that potential links via smaller clouds are not missed.

\subsubsection{Supercloud identification}

The four most distinctly identifiable superclouds in our sample are the turquoise, red, blue, and pink cloud groups, showing the clearest alignment in their orientation and 3D distribution in Fig. \ref{fig:superclouds_general}. The red group corresponds to the Radcliffe Wave, first reported by \citet{alves2020}. Our results are in close agreement with their analysis, with a total mass of $3.6 \times 10^6$ M$_\odot$ and a pitch angle of 31.4$\degr$. 
The blue group corresponds to the Split \citep{Lallement2019}. 
We introduce the name Natrix\footnote{The three leftmost superclouds are named after genera of snakes. For the turquoise group, the name Natrix comes from Natrix natrix, the grass snake, which is the most common snake in Austria. The orange cloud group is named Malpolon after Malpolon monspessulanus, the largest snake found in Portugal. Finally, Anguis comes from Anguis fragilis, the common slow worm, which resembles a snake in appearance but is actually a type of legless lizard. The names link the two star-forming superclouds with true snakes and the one without any known SFRs with the mimic.} Cloud for the turquoise cloud group. 
Lastly, the pink group is interpreted as an extension of the Sagittarius Spur \citep{kuhn2021}, which we name the Sagittarius Spur Extension (SSE). 

For the more disrupted orange and yellow cloud group, we adopt the name Malpolon Cloud. Its irregular morphology and puffed-up appearance in Z is likely caused by the influence of nearby high-mass stars, as traced by \cite{zari2023} and the ALS III Catalog of Galactic OB stars \citep{pantaleoni2025}, which is a recent expansion and \textit{Gaia} DR3 update of the catalog in \cite{ALSII}.

The green supercloud, which we name the Vela Ridge Cloud, is the least spatially coherent among the identified superclouds, with its components fragmented and widely separated. 
A possible explanation is that the Vela Ridge Cloud itself is not an independent supercloud but rather a disrupted part of the SSE, as orbital tracebacks of young star clusters hint that past clustered supernovae may have occurred in this region (Swiggum et al., in prep).
This theory of a past connection is also supported by the presence of high-mass stars in the canyon that separates clouds 4 and 33 from cloud 2. The stellar component extends the Spur further to almost double the initially defined length, linking the separated cloud components \citep{pantaleoni2025}. 
Spatially, it is not unlikely that clouds 29 and 39 (referred to as the L-clouds) are being sheared off from the SSE due to external forces, such as feedback from high-mass stars in this region \citep{zari2023, pantaleoni2025}, or supernovae from the Messier 6 cluster family that partially occupies this region \citep{swiggum_most_2024}. Additionally, clouds 23 and 26 may have originally been part of the Split, which is supported by their physical connection. The displacement is likely a result of the event responsible for the ring they surround. We kept the Vela Ridge Cloud as a supercloud in this work, as we cannot verify any of the alternative interpretations yet.

Lastly, the brown supercloud, located at the upper edge of the map, is given the name Anguis Cloud. While the two supercloud components are not connected, comparing our results with the extended version of the \citetalias{Edenhofer2024} dust map and the dust map by \cite{vergely2022} shows that they are likely part of a supercloud extending beyond the map. 

\subsubsection{Supercloud properties}

All superclouds we identify are truncated by the artificial boundary of the 3D reconstruction and are therefore incomplete. Their cloud components are either cut off at the border of the map or show potential extensions when compared to the extended 2 kpc \citetalias{Edenhofer2024} dust map and the dust map from \cite{vergely2022}. These maps are not used for the segmentation itself, due to their decreasing reliability beyond the boundary of the main map and their lower resolution. Instead, they serve as additional references for assumptions on the possible continuation of the superclouds. 

As shown by \cite{alves2020}, the Radcliffe Wave extends to the Cygnus X complex. \cite{zucker2020} calculated accurate distances for SFRs in the Star Formation Handbook \citep{reipurth2008_sfr_north, reipurth2008_sfr_south} up to 2.5 kpc from the Sun. Comparing their results with the superclouds, we find that the Natrix Cloud is likely to extend to the Rosette Nebula, while the SSE connects to M20, Lagoon, L291, NGC6604, M17, M16, and L379. It is interesting to note that each supercloud, excluding the Anguis Cloud, contains at least a few known SFRs. This absence could indicate a less active or earlier phase of cloud evolution, but it might also result from observational restraints or incomplete coverage of the region.

The masses of the inferred superclouds range from $1.1$ to $3.6 \times 10^6$~M$_\odot$, with the Anguis Cloud as an outlier at $7.8 \times 10^5$~M$_\odot$. All superclouds, except the Anguis Cloud, show lengths ranging from approximately 1.5 to 2.5~kpc, which is calculated using PCA and taking all cloud components into account. A selection of the calculated cloud properties, including mass and orientation, can
be found in Table \ref{tab:supercloud_properties}. Their mean orientation is characterized by a face-on pitch angle $\alpha = 33.5 \pm 4.0 \degr$, and edge-on angle $\beta = 87.7 \pm 4.2 \degr$, making an association with a major spiral arm unlikely \citep{Franco2002, efremov2009}. We show the face-on orientation of the superclouds in the right panel of Fig. \ref{fig:superclouds_general}. The identified superclouds are consistent with the characteristic scales and masses predicted in earlier studies \citep{Elmegreen1987, park2023}, representing the largest scale of ISM structures in the local Milky Way and likely throughout the Galaxy as a whole. 

In total, the \citetalias{Edenhofer2024} dust map has a gas mass of $3.3 \times 10^7$ M$_\odot$. For Z = $\pm \, 100$ pc, the volume filling factor of the superclouds is 9.94\%, containing 55.95\% of the gas mass in the Galactic plane.

\subsubsection{Undulation pattern}

Notably, the undulation observed in the Radcliffe Wave is not unique. All but one of the seven superclouds show an undulating pattern in the Z/Y plane, indicated by the color pattern of Fig.~\ref{fig:structures_cc_meanz}, and are likely oscillating \citep{ konietzka2024}. Here, the upper part of the Radcliffe Wave appears flat in Z, as a result of averaging the change in height across the entire length of HOP cloud 0, erasing the underlying pattern. 

\begin{figure}[!t]
        \centering
        \includegraphics[width = 0.5\textwidth]{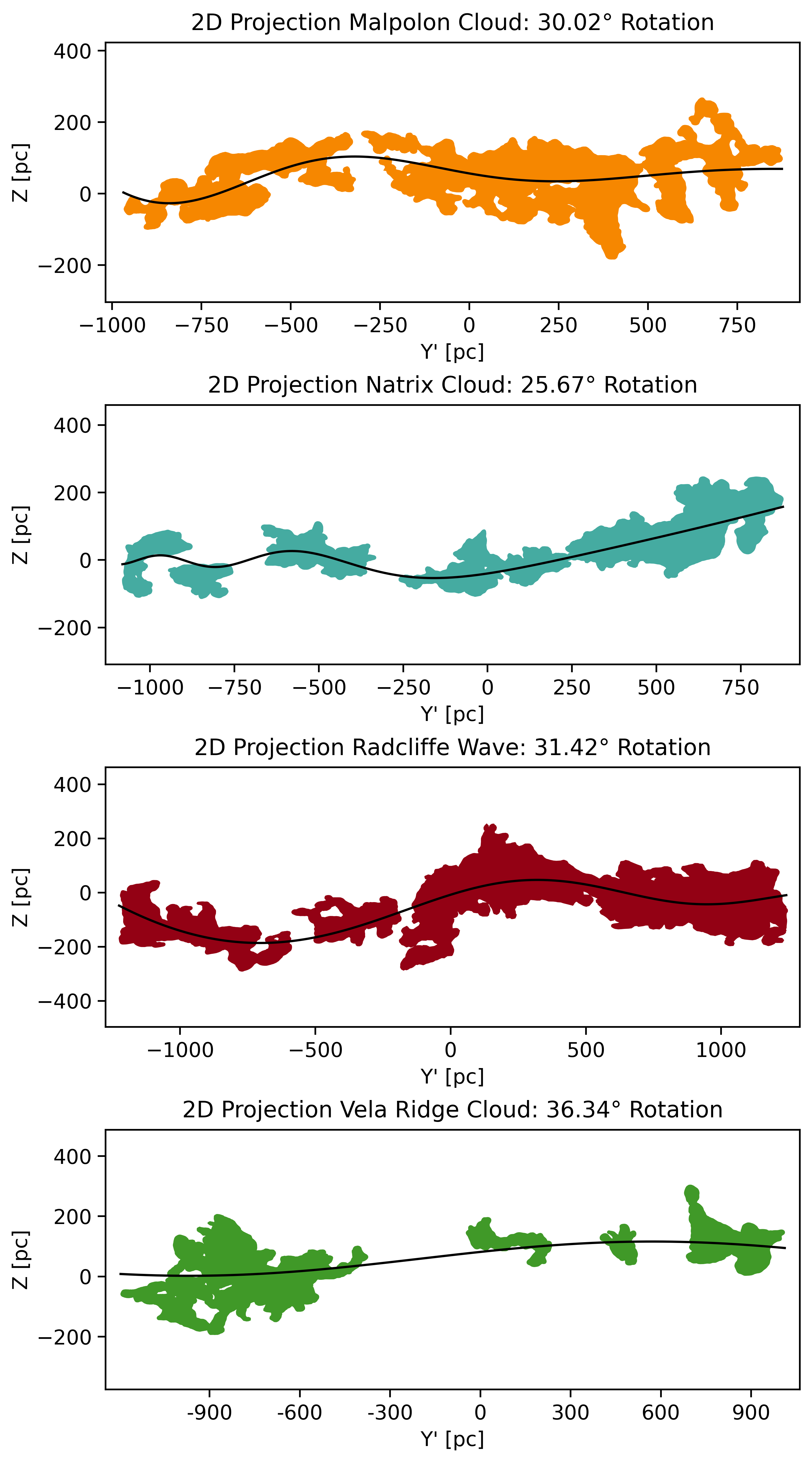}
        \caption{2D projection in the Z/Y$^\prime$ plane of the selected superclouds with a sinusoidal fit. For the fit, the clouds are rotated counterclockwise by their face-on angle ($\alpha$) in the X/Y plane. The first three clouds are fit with a damped sinusoidal, the fourth cloud with a simple sinus function. All fit parameters are listed in Table \ref{tab:fit_parameters}.
        }
        \label{fig:superclouds_undulation}
    \end{figure}

In Fig.~\ref{fig:superclouds_undulation} we show the sinusoidal fit for the selected superclouds: the Malpolon Cloud, Natrix Cloud, Radcliffe Wave, and Vela Ridge Cloud. The SSE and the Anguis Cloud are excluded from the analysis, as they show a vertical displacement, but it is not as prominent as for the previous examples. The Split is also excluded, as it appears unusually flat, with only two noticeable downward extensions at Y = -200 pc (referred to as the "C" and discussed in \citealt{edenhofer2024_c}) and Y = 500 pc. The three superclouds excluded from the fitting can be seen in Fig. \ref{fig:superclouds_flat}. All resulting fit parameters are listed in Table~\ref{tab:fit_parameters}, the fitting approach is simplified to give an initial characterization of the underlying undulation.

For the Malpolon Cloud, we excluded the three highest altitude clouds (clouds 10, 16, and 17) while fitting. It is likely that these clouds were pushed up by nearby high-mass stars  \citep{zari2023, pantaleoni2025}, and therefore distort the inherent structure of the supercloud. The Malpolon Cloud, Natrix Cloud, and Radcliffe Wave are well described by a damped sinusoidal wave. For the Malpolon Cloud, the parameters indicate decay in the amplitude and growth in the wavelength. Both damping parameters for the Natrix Cloud indicate growth over distance. For the Radcliffe Wave, the parameters show decay in amplitude and wavelength. 

The Vela Ridge Cloud is fitted with a simple sinusoidal function. While the undulation is visible, the large wavelength is likely a consequence of the fragmented nature of the supercloud. This, together with the incompleteness, supports the idea that Vela Ridge is not an individual supercloud. 

\section{Discussion}

A key implication of the main results in this paper is that these superclouds are unlikely to be associated with the spiral arms of the Milky Way, which have significantly smaller pitch angles (on average 10$^\circ$, as shown in \citealt{reid2019}). This includes the most massive of the local superclouds, the Radcliffe Wave, which is normally associated with the Local Arm (also known as the Orion Arm or the Orion Spur; see, e.g., \cite{swiggum2022}). Instead, the local superclouds seem to be part of a distinct, coherent structure that may reflect larger-scale dynamical processes in the Milky Way, perhaps akin to the high-pitch-angle gas lanes observed in other galaxies. 

Additionally, given that none of the seven superclouds seem to be associated with a Milky Way spiral arm, this implies that there is no obvious external trigger for the formation of GMCs inside of them, like a spiral shock. An initial comparison with the spatial distribution of known star-forming GMCs shows that nearly all lie within these superclouds, with many concentrated along their central regions of higher density, a result aligning with earlier results from \cite{Dame1986} and \cite{Elmegreen1987}. This supports a molecular cloud formation scenario in the local Milky Way in which H$_2$ forms in the denser regions of the neutral atomic cold medium without an obvious external trigger mechanism. However, this interpretation is tentative, and more work is needed to qualify this statement. Since GMCs are the primary sites of star formation, the superclouds presented in this work are likely to play an indirect but critical role in shaping the star formation process in the local Milky Way. 

Of the seven superclouds identified in this study, only two were previously known: the Split \citep{Lallement2019} and the Radcliffe Wave \citep{alves2020}. While the Split lies on the Galactic midplane, the Radcliffe Wave surprisingly undulates above and below it. Recent work by \cite{konietzka2024}, using young stellar clusters, confirmed that the undulations of the Radcliffe Wave represent vertical oscillations relative to the plane. Our results show that such undulations are not unique to the Radcliffe Wave. In fact, all superclouds, except the Split, exhibit similar vertical displacements. The origin of these undulations remains uncertain. One possibility is that they resulted from shocks or gravitational perturbations as superclouds interacted with spiral arms in the past, but further investigation is needed to test this scenario.

The identified superclouds can be associated with multiple known structures in the solar neighborhood. 
Spatially, the Anguis and the Malpolon Cloud appear to be a small section of the Cepheus Spur \citep{ALSII}. This spur extends from Cygnus OB2 toward the Galactic anticenter, as traced by young stellar populations and in more extended dust maps \citep[e.g.,][]{Lallement2019}. 
The bottom part of the Radcliffe Wave (clouds 27, 34, 30, 12, and 18) and of the Vela Ridge Cloud (clouds 21 and 2) define the boundaries of the GSH 238+00+09 supershell \citep{heiles1998}. This kiloparsec-long dust shell likely formed from clustered supernovae associated with the Cr135 cluster family \citep{swiggum_most_2024}, and occupies the full width between the two superclouds.
Recently, \cite{soler2025} calculated the kinetic energy densities of the local ISM based on the \citetalias{Edenhofer2024} dust map. Kinetic energy overdensities represent areas that differ roughly by a factor of ten or more from the mean value. The largest of these overdensities can be connected to the Radcliffe Wave and the Split, likely as a result of past supernovae or large-scale galactic dynamics.

\begin{figure}
        \centering
        \includegraphics[width = 0.5\textwidth]{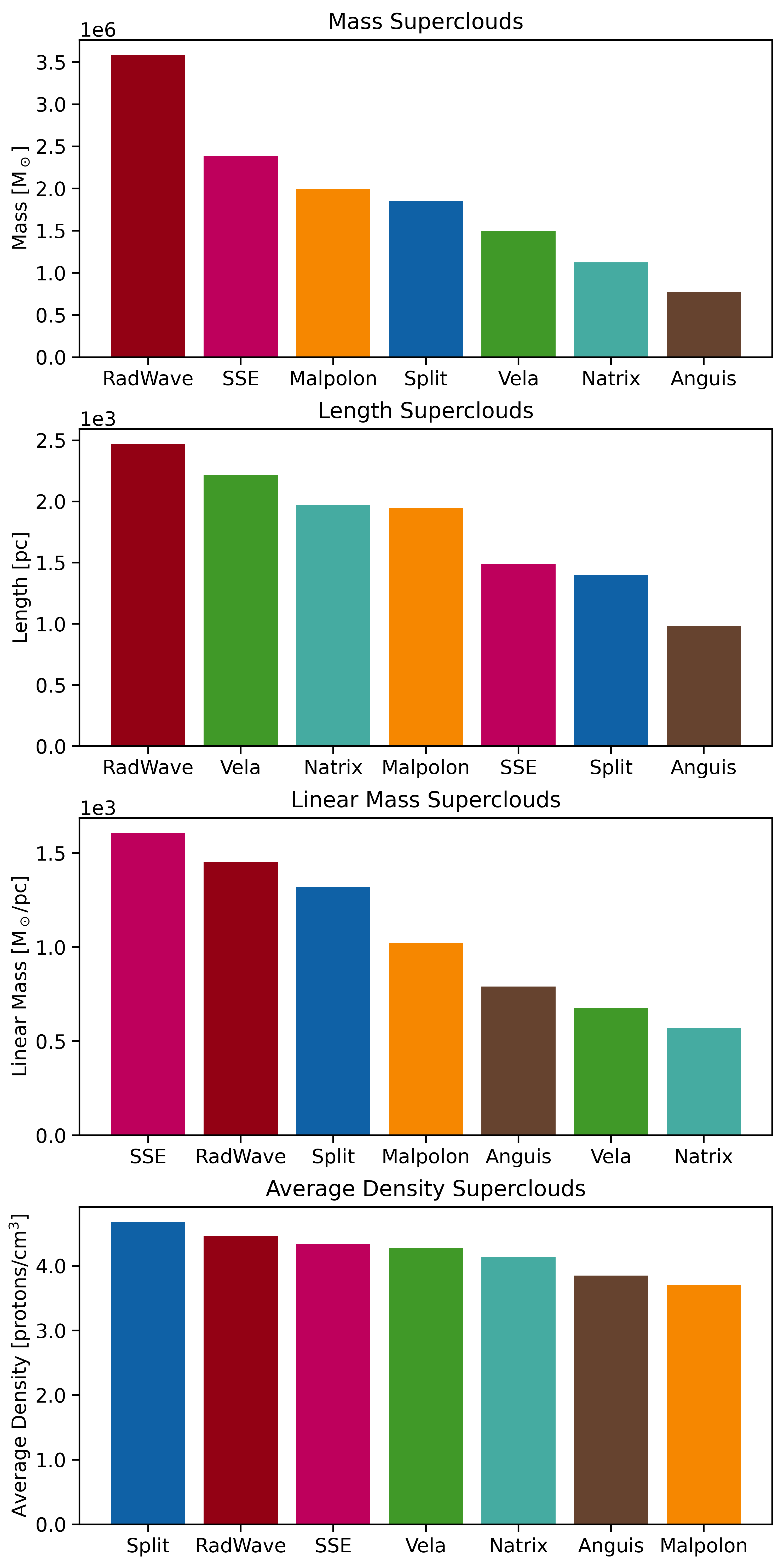}
        \caption{Properties calculated for the seven superclouds. The colors have the same meaning as in Fig. \ref{fig:superclouds_general}. RadWave refers to the Radcliffe Wave. Top: Total mass of the superclouds. Second: Total length of the superclouds. Third: Linear mass of the superclouds, the ratio of total mass to total length. Bottom: Average density of the superclouds. 
        The trends represent a lower bound, as all superclouds are incomplete and likely continue beyond the map.}
        \label{fig:superclouds_properties}
    \end{figure}

Figures~\ref{fig:superclouds_undulation} and \ref{fig:superclouds_properties} reveal an intriguing clue about the stability of local superclouds. While their mass-per-length varies by factors of 3–4, their average densities differ by only about 10\%. This points to a state of approximate pressure equilibrium between the superclouds and the surrounding medium. In such a scenario, external pressure sets the density scale, while individual superclouds adjust their masses and sizes within this constraint, resulting in structurally diverse clouds with similar internal conditions. In other words, although their column densities (and thus gravitational binding) differ, their internal pressures remain roughly comparable. We propose that these structures self-regulate their physical sizes and internal kinematics to maintain a pressure balance with their environment. This would naturally explain the coexistence of superclouds with varying masses and dimensions within a common large-scale ISM context, consistent with observations.

The well-defined, quasi-parallel pattern of superclouds shown in Fig.~\ref{fig:superclouds_general} suggests a relatively undisturbed local ISM. While stellar feedback is clearly reshaping the gas on kiloparsec scales, for example, in the case of the Heiles supershell \citep{swiggum_most_2024}, it does not appear to have significantly disrupted the overall supercloud pattern. Based on the formation scenario for GMCs as discussed above, we can make a conservative, lower age estimation by considering the ages of the oldest stars formed within the superclouds. For the Radcliffe Wave, this corresponds to approximately 20-30 Myr. Compared to kinematic timescales, this lower age estimate is short. Recent measurements of the local ISM velocity dispersion, approximately $11\,\mathrm{km\,s^{-1}}$ \citep[see][and references therein]{soler2025}, suggest that a perturbation would require over $90\,\mathrm{Myr}$ to cross the full length of a supercloud. This timescale is about three times longer than the estimate based on stellar ages and exceeds the typical lifetimes of CO clouds \citep[see, e.g.,][]{mivilledeschenes2017,chevance2020}. These discrepancies imply that superclouds are unlikely to form by a bottom-up process, i.e., from small to large scales, as would be expected in a scenario driven purely by stellar feedback. Instead, large-scale Galactic dynamics and instabilities probably play a central role in the formation of the kiloparsec-scale density structures we identify as superclouds \citep[see, e.g.,][]{kim2002,kim_ostriker_tigress2017,smith2020,colman2022}. This does not exclude the influence of energy and momentum injection from interior star formation \citep{swiggum_most_2024}, but it points to the need for additional physical processes to explain the observed coherence and scale of these structures.

\section{Conclusion}

The most significant outcome of this study is the discovery of a remarkably organized gas distribution within the local kiloparsec of the Milky Way. This distribution follows a distinct pattern of quasi-parallel gas lanes extending over several kiloparsec (see the right panel of Fig.~\ref{fig:superclouds_general}). These lanes contain substantial masses, on the order of 10$^6$ M$_\odot$, and exhibit pitch angles of approximately 30$^\circ$. Given their size, mass, and likely composition (predominantly atomic hydrogen), we refer to these structures as superclouds, following the terminology of \cite{elmegreen1983}. This identification has only become possible with the advent of large-scale 3D dust maps based on data from the \textit{Gaia} mission \citep{gaiaDR3}, which allow us to trace the local gas distribution in 3D with unprecedented detail. Unlike earlier studies of superclouds that relied on position-position-velocity analysis, these new maps offer a clearer, spatially resolved view of the ISM.

The discovery of a coherent network of superclouds within the local kiloparsec marks a step forward in our understanding of the Milky Way’s ISM. These massive, elongated structures appear to serve as scaffolds for molecular cloud formation and star formation on Galactic scales. Much remains to be understood: what these structures truly are, how they form, and how they give rise to GMCs and, ultimately, stars. Their morphology, stability, and apparent independence from the traditional spiral arm framework raise important questions about the large-scale dynamics of the Galaxy. As new data and analysis tools continue to improve our view of the 3D ISM, these findings offer a foundation for future investigations into the interplay between galactic structure, gas dynamics, and star formation. As we refine our 3D view of the nearby Galaxy, we are beginning to define a solid benchmark for connecting the complex structure of the Milky Way with the broader context of galaxy evolution and star formation in the Universe.


\begin{acknowledgements}
The authors are particularly grateful to Juan Soler for exciting discussions on the superclouds, for suggesting the HOP algorithm, and for providing helpful comments on the discussion section. We thank Bruce and Debra Elmegreen for insightful discussions on the nature of the superclouds.  We also thank Tine Colman for helpful comments on the HOP algorithm, and Alvaro Hacar, Robert Benjamin, and Alyssa Goodman for valuable feedback on the physical state of superclouds. L.A.K. is grateful to Moa Huppenkothen for helpful discussions on the naming of the superclouds.
Co-funded by the European Union (ERC, ISM-FLOW, 101055318). Views and opinions expressed are, however, those of the author(s) only and do not necessarily reflect those of the European Union or the European Research Council. Neither the European Union nor the granting authority can be held responsible for them. The results in this paper were based on observations obtained with \textit{Gaia}, an ESA science mission with instruments and contributions directly funded by ESA Member States, NASA, and Canada. 
This research has made use of NASA's Astrophysics Data System, as well as publicly available software libraries, including Astropy \citep{robitaille_astropy_2013}, NumPy \citep{numpy}, scikit-learn \citep{scikit-learn}, matplotlib \citep{matplotlib}, and plotly \citep{plotly}. The computational results have been achieved using the Austrian Scientific Computing (ASC) infrastructure.
\end{acknowledgements}

\bibliographystyle{aa} 
\bibliography{ref.bib} 

\appendix
\section{Rejected segmentation methods}
\label{ap:methods}

Here, the additional three segmentation methods are briefly discussed, along with the reasons for their rejection.

    \subsection{Scikit-Image}
    
    Scikit-Image \citep{scikitimage} is a tool for processing and analyzing 3D datasets, making it possible to extract higher density features in the dust data and labeling connected regions. It offers a fast and straightforward segmentation approach, but has a few limitations that make it unsuitable for this work. First, the method requires the minimum density threshold, the only parameter the segmentation depends on, to be predefined. The data are then segmented by distinguishing between high- and low-density voxels, depending on the given threshold. Although the labeling option makes it possible to identify connected structures, the approach is too simple to allow for any fine-tuning.
    
    \subsection{Dendrograms}
    
    Astrodendro \citep{astrodendro} is a package for constructing dendrograms; hierarchical structures used to identify and analyze nested features. While dendrograms are effective for analyzing and identifying hierarchical substructures, the size of the dust map makes it difficult to analyze regions of interest, due to the large number of resulting branches. This can be avoided by fine-tuning the parameters, but the more effective way is to extract smaller regions for the analysis. However, turning the map into smaller segments is not the preferred approach.

    \subsection{Convolutional neural network}
    
    The convolutional neural network (CNN) approach is based on \cite{kim2020}, who developed the algorithm for unsupervised image segmentation in 2D. It was later modified at the DS/Univie (private communication with Amanda de Paula Cristino Hirschl) for 3D applications. Originally adapted to find gaps in the dust map by \cite{Leike2020}, it can also be applied to detect high-density regions. In this approach, the CNN processes the dust data by extracting structural features based on density. This is done through convolutional layers, which enable the network to learn spatial patterns on different scales. The output is a binary dataset that indicates whether a data point belongs to a structure or is rejected as part of the background. The CNN enables a data-driven segmentation that does not depend on an external density threshold definition, but is not suitable to further segment the defined large-scale structures. This limitation makes it difficult to properly segment the dust map.

\section{Cloud numbering}

\begin{figure}[!ht]
        \centering
        \includegraphics[width = 0.48\textwidth]{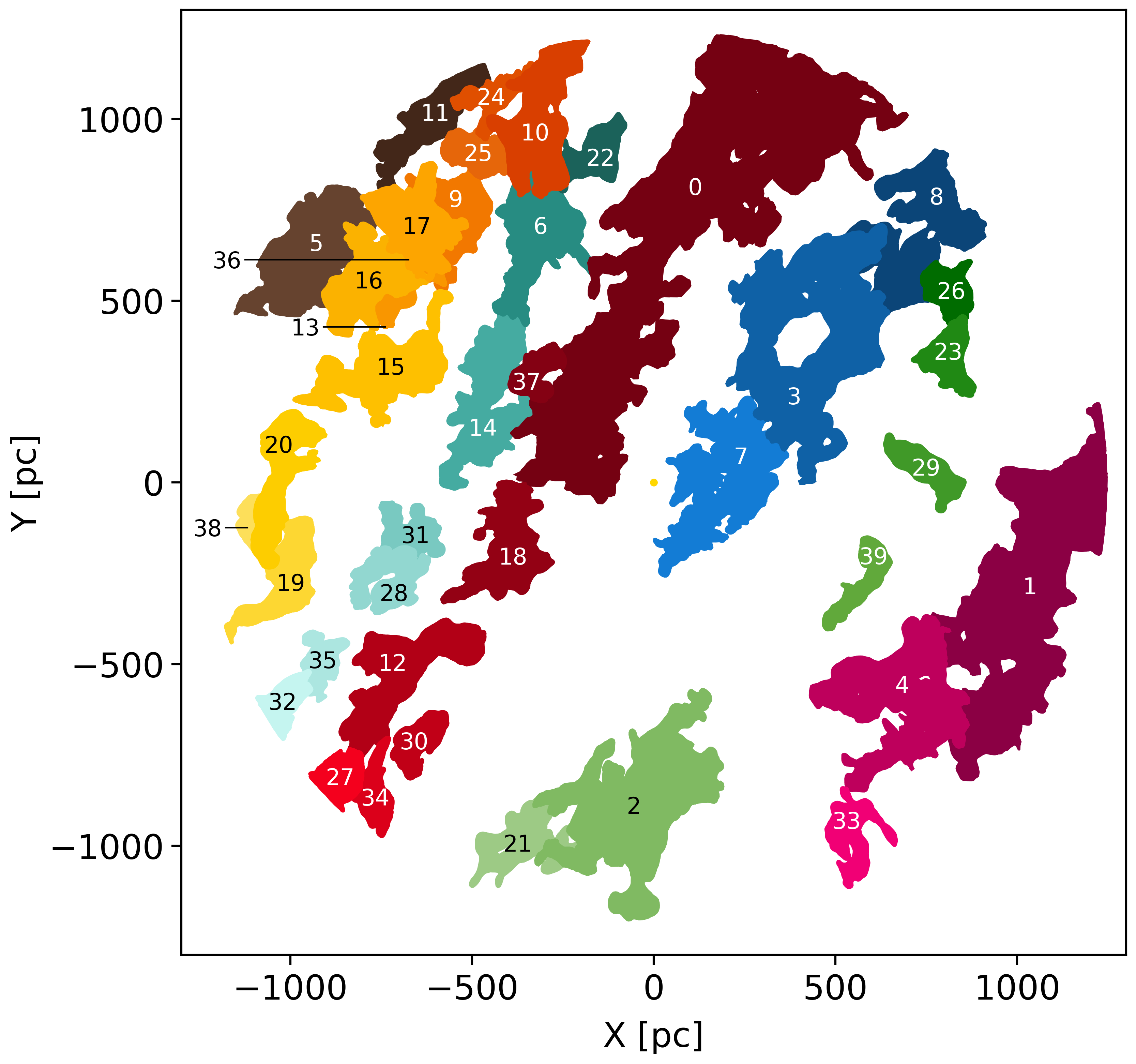}
        \caption{Same as Fig. \ref{fig:structures_general} (left) but with additional cloud ID numbers.}
        \label{fig:structures_full_with_id}
    \end{figure}

\begin{figure}[!ht]
        \centering
        \includegraphics[width = 0.48\textwidth]{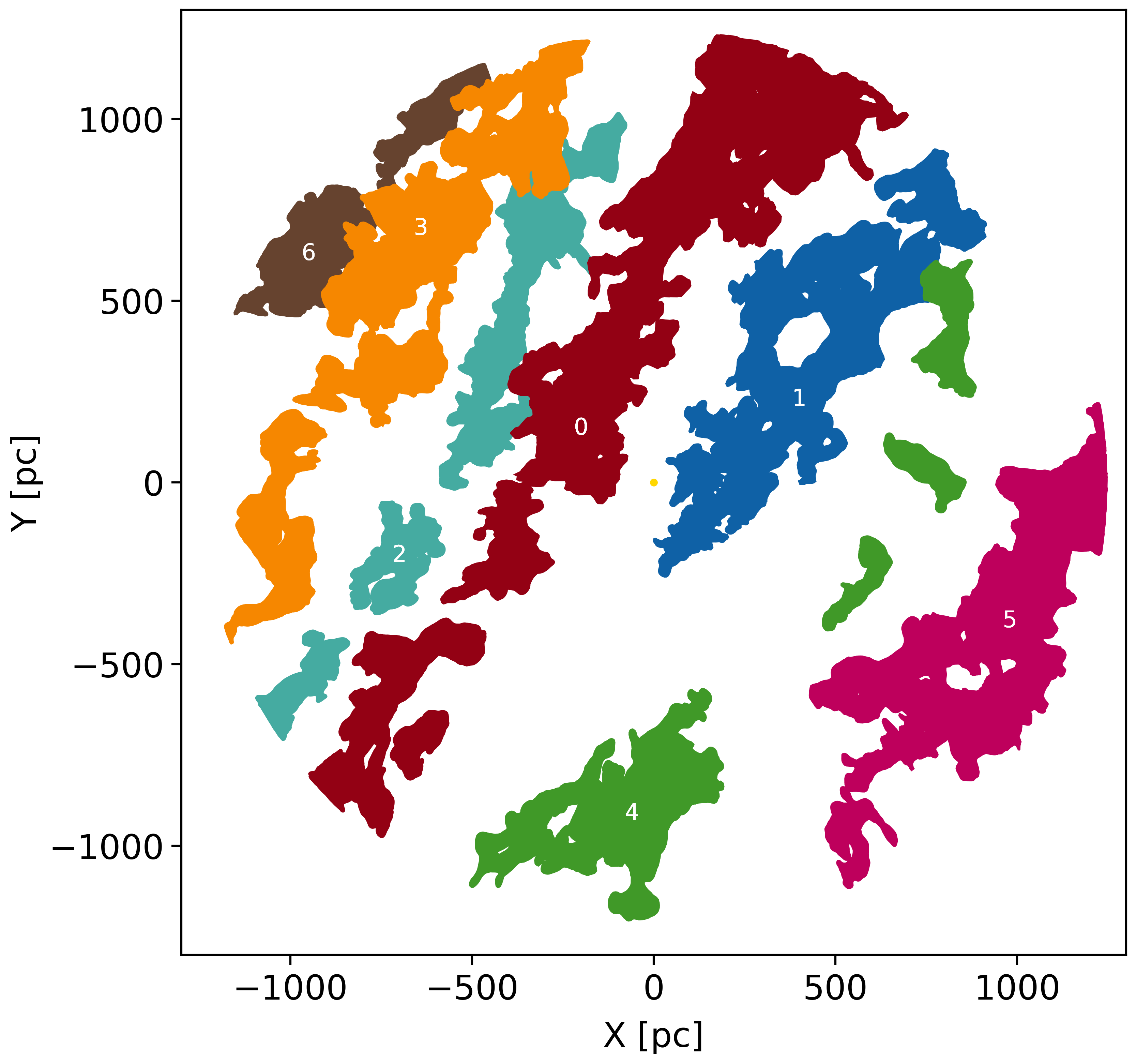}
        \caption{Same as Fig. \ref{fig:superclouds_general} (left) but with additional supercloud ID numbers.}
        \label{fig:superclouds_full_with_id}
    \end{figure}

Each cloud and supercloud is assigned a number for identification purposes. The HOP cloud numbers are assigned in Fig. \ref{fig:structures_full_with_id}, the supercloud numbers in Fig. \ref{fig:superclouds_full_with_id}. Table \ref{tab:superclouds_table} indicates which HOP clouds are part of each supercloud.

    \begin{table}[!h]
        \centering
        \vspace{0.16cm}
        \caption{HOP cloud components for each supercloud; the ID refers to the cloud ID as shown in Fig. \ref{fig:structures_full_with_id}.}
        \begin{tabular}{l l l}
            \hline
            ID & Supercloud & HOP Cloud\\
            \hline \hline
            0 & Radcliffe Wave & 0, 37, 18, 12, 30, 27, 34 \\
            1 & Split & 8, 3, 7 \\
            2 & Natrix Cloud & 22, 6, 14, 31, 28, 35, 32 \\
            3 & Malpolon Cloud &  24, 10, 25, 9, 17, 36, 16 \\
            & & 13, 15, 20, 38, 19 \\
            4 & Vela Ridge Cloud & 26, 23, 29, 39, 2, 21 \\
            5 & Sagittarius Spur Extension & 1, 4, 33  \\
            6 & Anguis Cloud & 11, 5\\
            \hline
        \end{tabular}
        \label{tab:superclouds_table}
    \end{table}

\section{Cloud properties}
\label{ap:cloud_properties}

Multiple properties are calculated for the HOP clouds as described in Sect. \ref{sec:methods} and can be found in Table \ref{tab:cloud_properties}. The supercloud properties are shown in Table \ref{tab:supercloud_properties}. The resulting fit parameters for the Malpolon Cloud, Natrix Cloud, Radcliffe Wave, and the Vela Ridge Cloud are shown in Table \ref{tab:fit_parameters}.

\begin{table*}[!ht]
    \centering
    \caption{Selection of parameters calculated for the 40 HOP clouds.}
    \begin{tabular*}{\textwidth}{@{\extracolsep{\fill}}l l r r r r r r r r}
    \hline
        $\, $Cloud ID & Supercloud & Cell Count & Mass [M$_\odot$] & L [pc] & X [pc] & Y [pc] & Z [pc] &$\alpha$ [$\degr$] & $\beta$ [$\degr$] \\  \hline \hline
        $\ \,$0 & Radcliffe Wave & 3142543 & 2.77e+06 & 1429.77 & 114.76 & 719.92 & -15.75 & 31.46 & 91.94 \\ 
        $\ \,$1 & Sag. Spur Extension & 2018035 & 1.81e+06 & 1100.18 & 1035.86 & -289.03 & -2.24 & 21.24 & 91.37 \\ 
        $\ \,$2 & Vela Ridge Cloud & 1168440 & 1.06e+06 & 654.03 & -54.58 & -893.01 & 18.71 & 31.39 & 101.79 \\ 
        $\ \,$3 & Split & 990721 & 9.39e+05 & 742.11 & 426.96 & 383.43 & 19.78 & 10.11 & 89.68 \\ 
        $\ \,$4 & Sag. Spur Extension & 693664 & 5.24e+05 & 466.71 & 683.74 & -580.29 & 39.30 & 298.92 & 144.88 \\ 
        $\ \,$5 & Anguis Cloud & 675154 & 5.05e+05 & 495.69 & -928.40 & 655.21 & 19.14 & 30.51 & 50.88 \\ 
        $\ \,$6 & Natrix Cloud & 569140 & 5.14e+05 & 543.46 & -311.49 & 722.62 & 60.55 & 10.18 & 78.85 \\ 
        $\ \,$7 & Split & 535072 & 5.43e+05 & 554.04 & 189.72 & 18.26 & 11.00 & 30.61 & 75.75 \\ 
        $\ \,$8 & Split & 474359 & 3.67e+05 & 466.99 & 727.94 & 683.38 & 0.83 & 30.37 & 89.98 \\ 
        $\ \,$9 & Malpolon Cloud & 415170 & 3.06e+05 & 413.61 & -564.91 & 726.98 & 35.75 & 13.46 & 133.17 \\ 
        $\ \,$10 & Malpolon Cloud & 376439 & 3.17e+05 & 461.37 & -306.69 & 999.70 & 243.40 & 9.10 & 102.14 \\ 
        $\ \,$11 & Anguis Cloud & 343470 & 2.71e+05 & 435.42 & -601.23 & 1014.20 & 33.62 & 41.08 & 37.51 \\ 
        $\ \,$12 & Radcliffe Wave & 338392 & 2.93e+05 & 504.32 & -698.04 & -529.46 & -179.72 & 43.88 & 112.12 \\ 
        $\ \,$13 & Malpolon Cloud & 318522 & 2.32e+05 & 333.37 & -723.67 & 566.25 & 54.48 & 25.75 & 68.03 \\ 
        $\ \,$14 & Natrix Cloud & 306527 & 2.47e+05 & 516.19 & -449.49 & 227.49 & -30.37 & 12.84 & 81.06 \\ 
        $\ \,$15 & Malpolon Cloud & 302052 & 2.19e+05 & 524.56 & -722.79 & 315.55 & 66.79 & 67.97 & 112.46 \\ 
        $\ \,$16 & Malpolon Cloud & 277775 & 1.64e+05 & 323.23 & -784.05 & 552.04 & 167.18 & 44.67 & 85.16 \\ 
        $\ \,$17 & Malpolon Cloud & 247255 & 1.66e+05 & 310.93 & -652.01 & 722.71 & 176.71 & 328.35 & 65.37 \\ 
        $\ \,$18 & Radcliffe Wave & 247032 & 2.82e+05 & 369.44 & -397.34 & -177.36 & -103.16 & 14.40 & 77.41 \\ 
        $\ \,$19 & Malpolon Cloud & 188997 & 1.40e+05 & 411.70 & -1019.06 & -254.26 & -23.77 & 27.12 & 88.54 \\ 
        $\ \,$20 & Malpolon Cloud & 188243 & 1.43e+05 & 454.32 & -1032.00 & 10.69 & 88.33 & 13.44 & 90.00 \\ 
        $\ \,$21 & Vela Ridge Cloud & 159910 & 1.09e+05 & 315.85 & -334.22 & -997.13 & -70.90 & 85.58 & 90.44 \\ 
        $\ \,$22 & Natrix Cloud & 159249 & 1.26e+05 & 279.32 & -166.65 & 909.66 & 153.50 & 48.65 & 131.09 \\ 
        $\ \,$23 & Vela Ridge Cloud & 145303 & 8.44e+04 & 283.46 & 819.62 & 335.96 & 137.91 & 345.86 & 137.22 \\ 
        $\ \,$24 & Malpolon Cloud & 138999 & 1.06e+05 & 319.05 & -422.34 & 1066.74 & 107.69 & 62.70 & 160.71 \\ 
        $\ \,$25 & Malpolon Cloud & 138713 & 1.13e+05 & 295.11 & -453.36 & 903.42 & 47.68 & 281.21 & 8.52 \\ 
        $\ \,$26 & Vela Ridge Cloud & 133466 & 9.65e+04 & 220.03 & 818.50 & 523.70 & 86.36 & 310.51 & 147.82 \\ 
        $\ \,$27 & Radcliffe Wave & 115059 & 8.50e+04 & 213.97 & -862.93 & -814.00 & -66.09 & 68.35 & 166.31 \\ 
        $\ \,$28 & Natrix Cloud & 104836 & 8.50e+04 & 266.80 & -725.34 & -272.38 & 34.74 & 56.11 & 113.74 \\ 
        $\ \,$29 & Vela Ridge Cloud & 104127 & 1.02e+05 & 262.17 & 749.41 & 37.72 & 95.62 & 306.39 & 93.83 \\ 
        $\ \,$30 & Radcliffe Wave & 95155 & 6.65e+04 & 231.40 & -638.95 & -714.05 & -131.42 & 37.42 & 114.98 \\ 
        $\ \,$31 & Natrix Cloud & 85936 & 5.87e+04 & 236.84 & -684.91 & -145.18 & -5.22 & 81.38 & 33.35 \\ 
        $\ \,$32 & Natrix Cloud & 84333 & 5.50e+04 & 236.98 & -1021.12 & -613.86 & 14.61 & 32.79 & 36.12 \\ 
        $\ \,$33 & Sag. Spur Extension & 72527 & 5.22e+04 & 271.71 & 546.04 & -964.51 & 126.08 & 1.19 & 80.83 \\ 
        $\ \,$34 & Radcliffe Wave & 67317 & 4.77e+04 & 275.54 & -766.10 & -860.39 & -154.85 & 355.28 & 92.10 \\ 
        $\ \,$35 & Natrix Cloud & 63846 & 3.77e+04 & 192.22 & -915.17 & -497.46 & -51.00 & 17.88 & 96.27 \\ 
        $\ \,$36 & Malpolon Cloud & 62484 & 4.58e+04 & 220.50 & -664.41 & 612.74 & -34.04 & 27.61 & 71.41 \\ 
        $\ \,$37 & Radcliffe Wave & 61628 & 4.13e+04 & 176.90 & -324.72 & 308.98 & 82.87 & 45.88 & 100.38 \\ 
        $\ \,$38 & Malpolon Cloud & 61550 & 3.85e+04 & 176.38 & -1097.81 & -124.34 & 0.56 & 345.45 & 109.98 \\ 
        $\ \,$39 & Vela Ridge Cloud & 60876 & 4.64e+04 & 282.88 & 565.15 & -277.92 & 113.30 & 32.39 & 103.70 \\ 
        \hline
    \end{tabular*}
    \label{tab:cloud_properties}
    \tablefoot{Each cloud is assigned a cloud ID and has listed which supercloud it is part of. The column cell count indicates the voxels per structure, L gives the cloud length calculated with PCA. The columns X, Y, and Z are the unweighted mean coordinates per cloud. The angles $\alpha$ and $\beta$ represent the face-on and edge-on angles of the clouds.}
\end{table*}

\clearpage
\begin{table*}[!t]
    \centering
    \caption{Selection of parameters calculated for the seven superclouds.}
    \begin{tabular*}{\textwidth}{@{\extracolsep{\fill}}l r r r r r r r r r}
    \hline
        $\ \, $Supercloud & Cell Count & Mass [M$_\odot$] & L [pc] & X [pc] & Y [pc] & Z [pc] & $\alpha$ [$\degr$] & $\beta$ [$\degr$] \\  \hline \hline
        $\ \,$Radcliffe Wave & 4067126 & 3.58e+06 & 2468.67 & -50.50 & 452.14 & -39.64 & 31.42 & 86.67 \\ 
        $\ \,$Split & 2000152 & 1.85e+06 & 1399.60 & 434.87 & 356.88 & 12.94 & 36.87 & 90.02 \\ 
        $\ \,$Natrix Cloud & 1373867 & 1.12e+06 & 1970.08 & -452.04 & 364.88 & 36.95 & 25.67 & 85.19 \\ 
        $\ \,$Malpolon Cloud & 2716199 & 1.99e+06 & 1946.38 & -660.98 & 568.46 & 97.84 & 30.02 & 82.99 \\ 
        $\ \,$Vela Ridge Cloud  & 1772122 & 1.50e+06 & 2215.88 & 126.15 & -619.12 & 33.26 & 36.34 & 85.31 \\ 
        $\ \,$Sagittarius Spur Extension & 2784226 & 2.39e+06 & 1487.87 & 935.37 & -379.19 & 11.45 & 35.85 & 95.50 \\ 
        $\ \,$Anguis Cloud & 1018624 & 7.76e+05 & 981.80 & -818.08 & 776.26 & 24.02 & 42.19 & 82.99 \\ \hline
    \end{tabular*}
    \label{tab:supercloud_properties}
    \tablefoot{Each supercloud is assigned a name. The column cell count indicates the voxels per structure, L gives the cloud length calculated with PCA. The columns X, Y, and Z are the unweighted mean coordinates per cloud. The angles $\alpha$ and $\beta$ represent the face-on and edge-on angles of the clouds.}
\end{table*}
\begin{table*}[t]
    \centering
    \caption{Fit parameters for the undulating superclouds. }
    \begin{tabular*}{\textwidth}{@{\extracolsep{\fill}}l r r r r r r}

    \hline
        $\, $ Supercloud & Amplitude $a$ [pc] & Wavelength $\lambda$ [pc] & Phase $\phi$ [rad] &Offset $d$ [pc] & Damp. $\epsilon_a$ [pc$^{-1}$] & Damp. $\epsilon_\omega$ [pc$^{-1}$]  \\
        \hline \hline
        $\ \,$Malpolon Cloud & 31.97 & 1137.40 &-3.10  & 58.15 & $-1.18 \times 10^{-3}$ & $-7.52 \times 10^{-5}$ \\
        $\ \,$Natrix Cloud & 70.53 & 1749.47 & -0.58 & -1.17 & $1.59 \times 10^{-3}$ & $-1.12 \times 10^{-3}$ \\
        $\ \,$Radcliffe Wave & -87.95 & 1748.00 & -3.09 & -13.01 & $-1.10 \times 10^{-3}$ & $2.88 \times 10^{-4}$ \\
        $\ \,$Vela Ridge Cloud & 56.90 & 3078.95 & 0.41 &  59.14 & & \\
        \hline
    \end{tabular*}
    \label{tab:fit_parameters}
    \tablefoot{The column $a$ denotes the amplitude, $\lambda$ the wavelength, $\phi$ the phase shift, and $d$ the vertical offset. The exponential damping parameters $\epsilon_a$ for the amplitude and $\epsilon_\omega$ for the frequency are only listed for clouds fitted with a damped sinusoidal, as described in Sect. \ref{methods:undulation_fitting}, instead of a standard sinusoidal function.}
\end{table*}

\FloatBarrier
\section{Smaller-scale clouds}
The final segmentation parameters were selected to identify large-scale structures within the dust map. Consequently, several smaller clouds that did not satisfy the final HOP criteria for minimum size or density threshold were excluded. While these clouds are not the focus of this study, they highlight the importance of studying smaller-scale structures alongside the superclouds. Complementary work on smaller-scale structures can be found in \cite{xie2024}. To focus on smaller-scale structures, we also performed several test runs with different parameter sets, revealing interesting features absent from the final segmentation.

One of the most prominent features when using a lower \texttt{mincells} value of 30.000 is a connection between the Malpolon Cloud (HOP cloud 15) and Natrix Cloud (HOP cloud 31). This roughly 200 pc long connection is visible by eye in the \citetalias{Edenhofer2024} dust map at approximately X = -730 pc and Y = 60 pc. The origin of this feature is currently unknown, a possible explanation is the formation through external influences such as feedback processes, or it may indicate that the two superclouds were more strongly connected in the past. Further investigation is needed to understand the origin and significance of this feature. 

Another cloud recovered in this segmentation is located below HOP cloud 20 at approximately X = -920 pc and Y = 120 pc. While this cloud does not appear to be associated with any of the superclouds, it is a prominent feature in the \citetalias{Edenhofer2024} dust map. Given its compact and isolated appearance, it may be of interest for studies focused on smaller-scale structures of higher density.

With the lower \texttt{mincells} value, we also recover the top of the Pegasus Ridge (Swiggum et al. in prep). This cloud lies below the Radcliffe Wave, with the top part appearing to wrap around the edge of the Local Bubble \citep{oneill2024}. As shown in Swiggum et al. (in prep), it further extends below the Radcliffe Wave, but the lower part is composed of material with densities too low to be recovered by HOP.

Reducing the \texttt{mincells} parameter further leads to the recovery of additional small-scale clouds. However, these features do not affect the overall large-scale supercloud structure and lie outside the scope of the present study.

\section{Superclouds without sinusoidal fits}

\begin{figure}[!h]
        \centering
        \includegraphics[width = 0.5\textwidth]{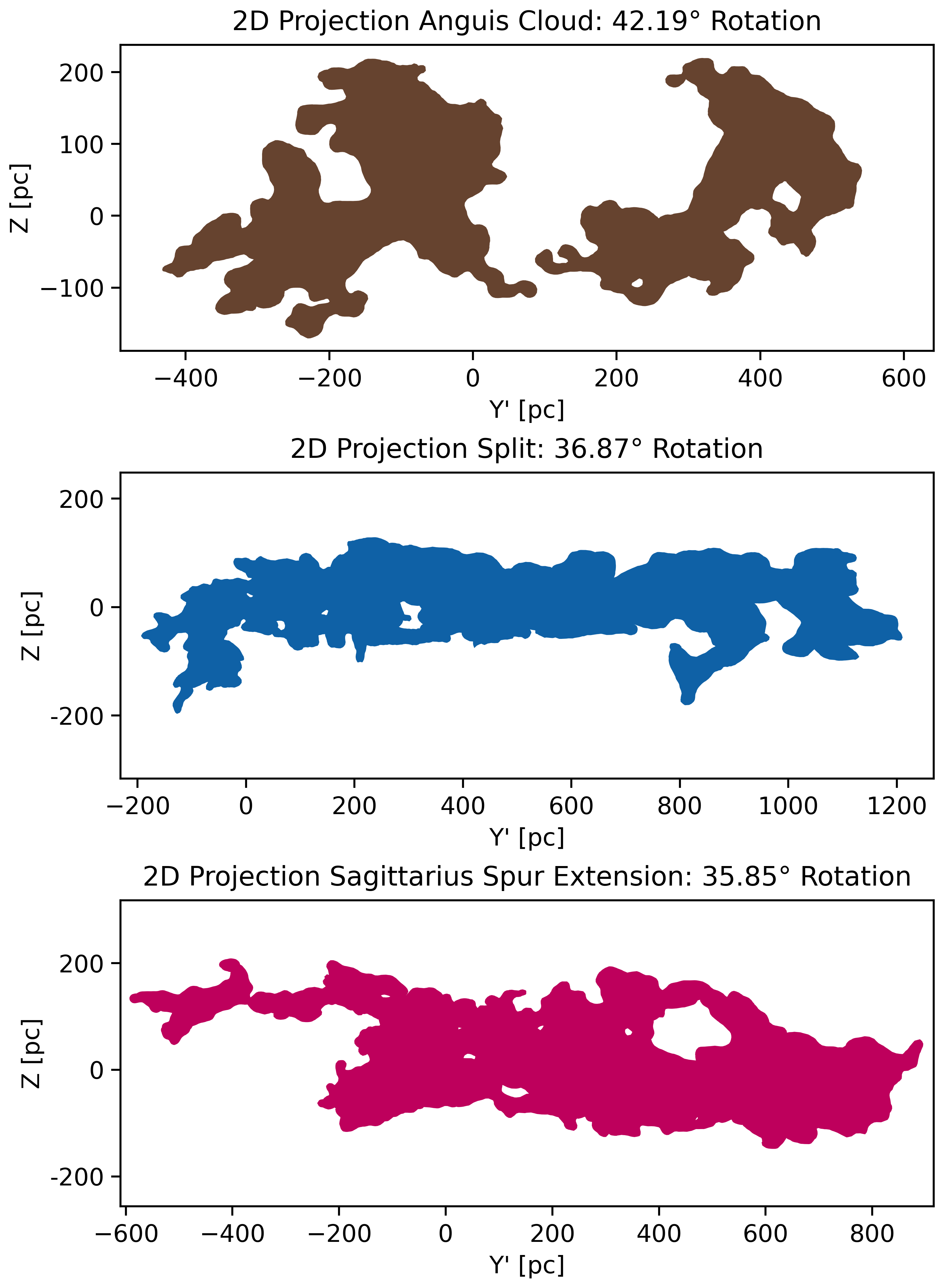}
        \caption{2D projection in the Z/Y$^\prime$ plane of the three superclouds without a sinusoidal fit. 
        }
        \label{fig:superclouds_flat}
    \end{figure}

The three remaining superclouds, which are not shown with a sinusoidal fit in Fig. \ref{fig:superclouds_undulation}, are shown in Fig. \ref{fig:superclouds_flat}. As mentioned, the SSE and the Anguis Cloud are excluded from the analysis, as the vertical displacement of the clouds is not as prominent as in the other examples. The Split is excluded, as it appears unusually flat, with only two noticeable downward extensions at Y = -200 pc (referred to as the "C" and discussed in \citealt{edenhofer2024_c}) and Y = 500 pc. 

\end{document}